\pdfoutput=1  
\documentclass[conference]{IEEEtran}

\usepackage[T1]{fontenc}
\usepackage[utf8]{inputenc}
\usepackage{microtype}
\usepackage{amsmath,amssymb}
\usepackage{graphicx}
\usepackage{booktabs}
\usepackage{array}
\usepackage{url}
\usepackage{xcolor}
\usepackage{cite}

\usepackage{tikz}
\usetikzlibrary{shapes.geometric, arrows.meta, positioning, fit, backgrounds, calc, decorations.pathreplacing}

\usepackage[hidelinks]{hyperref}

\tikzset{
  block/.style = {
    rectangle, rounded corners=3pt,
    draw=black!70, fill=white,
    text centered, font=\small,
    minimum height=1.6em, inner xsep=6pt, inner ysep=4pt
  },
  crdt/.style = {
    rectangle, rounded corners=3pt,
    draw=black!80, fill=gray!10,
    text centered, font=\small\bfseries,
    minimum width=2.0cm, minimum height=2.2em,
    inner xsep=4pt, inner ysep=4pt
  },
  role/.style = {
    font=\scriptsize\itshape, text=black!60
  },
  product/.style = {
    rectangle, rounded corners=3pt,
    draw=black!80, fill=gray!20,
    text centered, font=\small\bfseries,
    minimum height=1.8em, inner xsep=8pt, inner ysep=5pt
  },
  delta/.style = {
    rectangle, rounded corners=3pt,
    draw=black, fill=white,
    font=\small\bfseries,
    minimum height=1.8em, inner xsep=8pt, inner ysep=5pt
  },
  arrow/.style = {-{Stealth[length=4pt]}, thick, black!70},
  darrow/.style = {-{Stealth[length=4pt]}, thick, black},
}

\newcommand{\didcrdt}{\texttt{did:crdt}}
\newcommand{\CAI}{commutativity, associativity, and idempotence}

\newif\ifarxiv\arxivtrue   
\ifdefined\arxiv\arxivtrue\fi
\ifarxiv
  \makeatletter
  \def\ps@IEEEtitlepagestyle{\def\@oddfoot{\mycopyrightnotice}\def\@evenfoot{}}
  \def\mycopyrightnotice{%
    \begin{minipage}{\textwidth}\footnotesize
    \copyright~2026 IEEE. Personal use of this material is permitted. Permission
    from IEEE must be obtained for all other uses, in any current or future media,
    including reprinting/republishing this material for advertising or promotional
    purposes, creating new collective works, for resale or redistribution to
    servers or lists, or reuse of any copyrighted component of this work in other
    works.%
    \end{minipage}%
    \gdef\mycopyrightnotice{}}
  \makeatother
\fi

\begin{document}

\title{\texttt{did:crdt}: Coordination-Free Decentralised\\Identifiers via Signed CRDTs}

\IEEEoverridecommandlockouts
\author{
  \IEEEauthorblockN{Hugo O'Connor}
  \IEEEauthorblockA{Anuna Research\\ hugo@anuna.io}
  \and
  \IEEEauthorblockN{Claire Barnes}
  \IEEEauthorblockA{Anuna Research\\ claire@anuna.io}
  \thanks{Source code (MIT OR Apache-2.0): \href{https://codeberg.org/anuna/did-crdt}{codeberg.org/anuna/did-crdt}.}
}

\maketitle

\begin{abstract}
Existing Decentralised Identifier (DID) methods require coordination, an agreed
global order of operations, to update a DID document: blockchain-anchored methods
incur fees and latency; lightweight peer methods (\texttt{did:key},
\texttt{did:peer}) offer no update mechanism; and Sidetree methods still require
blockchain ordering for finality.
We present \didcrdt{}, a DID method that targets W3C DID Core and removes the need
for coordination entirely: there is no ledger, no sequencer, and no global total
order.
Each DID document is composed of signed Conflict-Free Replicated Data Types
(CRDTs), one per document field, each chosen so that concurrent edits merge
deterministically.
By the CALM Theorem, the state-merge path is then \emph{confluent}: replicas that
see the same updates reach the same document in any arrival order.
The signed-delta path needs only \emph{causal delivery}, applying an update after
those it builds on, which is far weaker than the total ordering ledgers impose and
needs no agreement protocol.
We are explicit about scope: every untrusted-peer path is authenticated, so
Byzantine fault tolerance (safety even when peers lie or send malformed data)
holds for signed deltas and verified-bundle replay, while the unauthenticated
state-merge path is a trusted-domain optimisation and key-compromise recovery is
bounded by revocation semantics.
We give the data and threat model, CRUD semantics, conflict resolution, and a
Rust reference implementation with property-based convergence tests and
microsecond-scale merge latency.
\end{abstract}


\section{Introduction}

Decentralised Identifiers (DIDs)~\cite{w3c-did-core} are the W3C standard for
self-sovereign identity (SSI), in which the subject, not an issuer or registry,
can control cryptographically verifiable identifiers. How fully a DID delivers
this is method-dependent: DID Core permits a controller distinct from the subject
and registries that are centralised, federated, or decentralised. A DID method
\emph{can}, however, place the identifier under its subject's sole control with no
issuer and no single point of revocation, which \didcrdt{} targets.
This property makes DIDs fundamental infrastructure for verifiable
credentials, decentralised authentication, and multi-device identity wallets.
Yet every production DID method accepts a damaging compromise: to support an
evolving document it coordinates through consensus, and to escape consensus it
freezes the document.

\textbf{The coordination trap.}
We use \emph{coordination} in the precise CALM sense~\cite{calm}: blocking
communication to reach \emph{agreement} on a global order before a node may act.
Such a total order is as hard as consensus (total-order broadcast is
consensus-equivalent~\cite{chandra-toueg}). Blockchain methods
(\texttt{did:ethr}~\cite{did-ethr}, \texttt{did:ion}~\cite{sidetree}) pay exactly
this cost: fees, seconds-to-minutes latency, and connectivity for every mutation.
Peer methods (\texttt{did:key}~\cite{did-key}, \texttt{did:peer}~\cite{did-peer})
avoid it by freezing the document, which then cannot rotate keys, add devices, or
synchronise. Sidetree~\cite{sidetree} adds a CRDT-like delta log but still anchors
it to a blockchain for total ordering before finality. No method achieves all four
of no consensus, offline mutation, cryptographic authorisation, and W3C DID Core
alignment; \didcrdt{} targets all four, realising the first three and substantially
the fourth (Section~\ref{sec:limitations}).

\textbf{Motivating example.}
Disaster response exposes the gap: responders authenticate today with centrally
issued credentials such as PIV-I/FRAC~\cite{cisa-cred-interop}, where a cached trust
anchor validates an existing badge offline, but \emph{enrolling} a responder or
\emph{revoking} a lost device needs the issuing authority a disaster removes (FEMA
puts revocation at a central update with an 18-hour window~\cite{fema-nims-cred}).
Consider instead one organisational DID co-managed by a few authorised devices at
incident command: offline and split across sites, each evolves the shared document
independently (registering radios, rotating a lost key, enrolling a replacement),
and when a channel reappears they exchange signed deltas and converge to
byte-identical state with no coordinator.

\textbf{Key insight.}
The CALM Theorem~\cite{calm} (Hellerstein and Alvaro) proves that any computation
expressible as a monotone function over a join-semilattice is confluent: replicas
converge without coordination. If every field of a DID document is modelled as a
CRDT~\cite{shapiro-crdts}, then each transition only adds information, their
composition is again a CRDT, and by CALM the merged state is confluent.
We map verification methods to a 2P-Set, services to an add-wins set, metadata
to a last-write-wins map, key rotation to a monotone sequence register, and
deactivation to an irreversible boolean latch (Section~\ref{sec:method}).
Signing each mutation at the application layer separates \emph{authorisation}
(local, cryptographic) from \emph{ordering}, so no consensus layer is needed.
The only ordering the protocol requires is causal delivery on the signed-delta
path; this, unlike a total order, needs no agreement protocol and remains
available under partition; by~\cite{mahajan-cac}, causal consistency is the
strongest model any always-available, partition-tolerant system admits.

\textbf{Contributions.}
(1)~A formal seven-field CRDT composition for the W3C DID document data model,
mapping each field to a CRDT type whose product satisfies \CAI.
(2)~The \didcrdt{} method specification: content-addressed identifier derivation
via BLAKE3-256, a typed delta vocabulary, and deterministic concurrent-merge
semantics.
(3)~A system and threat model and security analysis (genesis attacks, delta replay,
rotation races, state-size DoS, creation-flood sybil resistance, discovery-layer
availability, and the limits of key-compromise recovery), with Byzantine fault
tolerance established for every untrusted-peer path (signed deltas and
verified-bundle replay) and explicitly \emph{not} claimed for the trusted-replica
state-merge optimisation.
(4)~A Rust reference implementation (no \texttt{unsafe}, WASM-compatible) with
property-based convergence tests over randomised delta sequences.

\section{Background}
\label{sec:background}

\subsection{CRDTs and the CALM Theorem}

A CRDT's~\cite{shapiro-crdts} merge $\sqcup$ forms a join-semilattice: it is
commutative ($A \sqcup B = B \sqcup A$), associative
($(A \sqcup B) \sqcup C = A \sqcup (B \sqcup C)$), and idempotent
($A \sqcup A = A$), hereafter CAI. CAI suffices for \emph{strong eventual
consistency}: replicas receiving the same updates converge regardless of delivery
order or duplication, with no agreement protocol.

The Shapiro et al.\ catalogue~\cite{shapiro-crdts} enumerates the CRDT
primitives used in this work: a \textbf{G-Set} (grow-only set, merge = union),
an \textbf{OR-Set} using the ORSWOT optimisation~\cite{orswot} (add-wins, no
unbounded tombstones), an \textbf{LWW-Map} (last-write-wins per key under a total
timestamp order), a \textbf{Max-Register} (merge = max), and a \textbf{boolean
latch} (merge = logical OR).
Timestamps use \textbf{Hybrid Logical Clocks} (HLCs)~\cite{hlc}, a triple
$(\mathit{physical\_ms}, \mathit{counter}, \mathit{node\_id})$ that is
causally consistent and bounded in skew from wall time.
The $\mathit{node\_id}$ is derived from the signer's public key, a deterministic
tiebreaker bound to that key.
The \emph{product} of CRDTs is again a CRDT, a corollary of~\cite{shapiro-crdts}:
CAI holds componentwise.

The \textbf{CALM Theorem}~\cite{calm} states a distributed program has a
coordination-free, confluent implementation iff it is monotone. Crucially, CALM
asks only that \emph{merging} never discards information; it is fine for the
resolved document to stop showing a fact. We distinguish
\emph{lattice} monotonicity (merge only moves state upward in the semilattice),
which CALM requires and every \didcrdt{} field satisfies, from \emph{observable}
monotonicity (a fact, once visible, stays visible), which several fields
deliberately break: a revoked key disappears from the document, and a delta is
rejected if it precedes its authorising key. The latter is not needed for
convergence but is why the signed-delta admission path is order-sensitive.
Confluence is safety, not liveness: convergence still needs a transport that
eventually delivers every update. The state-merge path tolerates any order; the
signed-delta path needs only causal delivery, which is not coordination and is
strictly weaker than the consensus-equivalent total-order
broadcast~\cite{chandra-toueg} that ledgers require.
Kleppmann~\cite{kleppmann-byzantine} shows content-addressed hash graphs make
CRDTs Byzantine-fault-tolerant; \didcrdt{} applies this in its Merkle DAG.

\subsection{System Model}
\label{sec:sysmodel}

\textbf{Replicas and storage.}
There is no central server. Each \emph{replica} holds, for every DID it tracks,
the full product-CRDT state plus its delta log. The pure core is storage-agnostic
(state lives in memory, persisted by the embedder if at all), and any transport
that delivers signed deltas suffices: gossip, HTTP, or out-of-band media.

\textbf{Adversary.}
We assume a computationally bounded adversary that controls the network
(observe, delay, replay, reorder, drop) and may operate Byzantine replicas. It
cannot forge signatures or BLAKE3 pre-images. Honest replicas accept only
signed deltas authorised by a non-revoked key (Section~\ref{sec:delta}).

\textbf{Recency.}
Without a total order there is no global ``latest'': CALM guarantees convergence
among replicas that have seen the same deltas, not freshness. The monotone
\texttt{versionId} lets a verifier order two states by delta inclusion, but
\emph{absolute} recency needs a freshness response from a well-connected peer, the
deliberate cost of offline, coordination-free operation.

\section{Related Work}
\label{sec:related}

\texttt{did:key}~\cite{did-key} derives the identifier directly from the public
key, making the document static; \texttt{did:peer}~\cite{did-peer} is likewise
static. \texttt{did:ethr}~\cite{did-ethr} stores documents as Ethereum
smart-contract events, inheriting gas fees and ${\sim}12$\,s confirmation latency;
Satybaldy et al.~\cite{satybaldy-benchmarks} benchmark such methods
(\texttt{did:ethr} at 12.9\,s, \$0.066/op), motivating dropping the ledger.
Sidetree~\cite{sidetree} is the closest structural predecessor: both use a
content-addressed delta log where each operation is a typed mutation.
The critical divergence is ordering: Sidetree requires a blockchain anchor
to impose a total order, whereas \didcrdt{} selects CRDT types whose merge
functions are order-independent by construction.

\textbf{Ledgerless verifiable history.}
KERI~\cite{keri} and \texttt{did:webvh}~\cite{didwebvh} drop the ledger but keep a
single controller's \emph{totally ordered} log, treating concurrent multi-device
writes as \emph{equivocation} to detect rather than state to merge. \didcrdt{}
instead composes commuting CRDTs that converge offline with no coordinator or
canonical log; CRDTs have served identity \emph{data} (e.g.\ DIF Identity Hubs) but
not, to our knowledge, the DID document itself as a method (Ceramic's
\texttt{did:3}, despite a CRDT-like reputation, uses blockchain-anchored fork-choice
that picks one canonical branch, not a commutative merge).

\section{The \texttt{did:crdt} Method}
\label{sec:method}

\subsection{Identifier Derivation}

A \didcrdt{} identifier is content-addressed: the controller constructs a
\emph{genesis delta} (an \texttt{AddVerificationMethod} operation with a
logical timestamp of zero), and the DID is the lowercase hexadecimal encoding
of the BLAKE3-256 hash~\cite{blake3} of the compact JSON tuple
$(\tau_0, \mathit{op}_0, \mathit{signer\_key})$, i.e.\
\texttt{did:crdt:}$\langle\text{BLAKE3-256}(\mathrm{json}(\tau_0, \mathit{op}_0, \mathit{key}))\rangle$.
This tuple is hashed \emph{before} the DID exists; the real DID then forms the full
key identifier (e.g.\ \texttt{did:crdt:abc\#key-0}). Creation needs no network,
registry, or synchronisation, and a given genesis delta deterministically yields the
same DID.

\subsection{CRDT Field Composition}

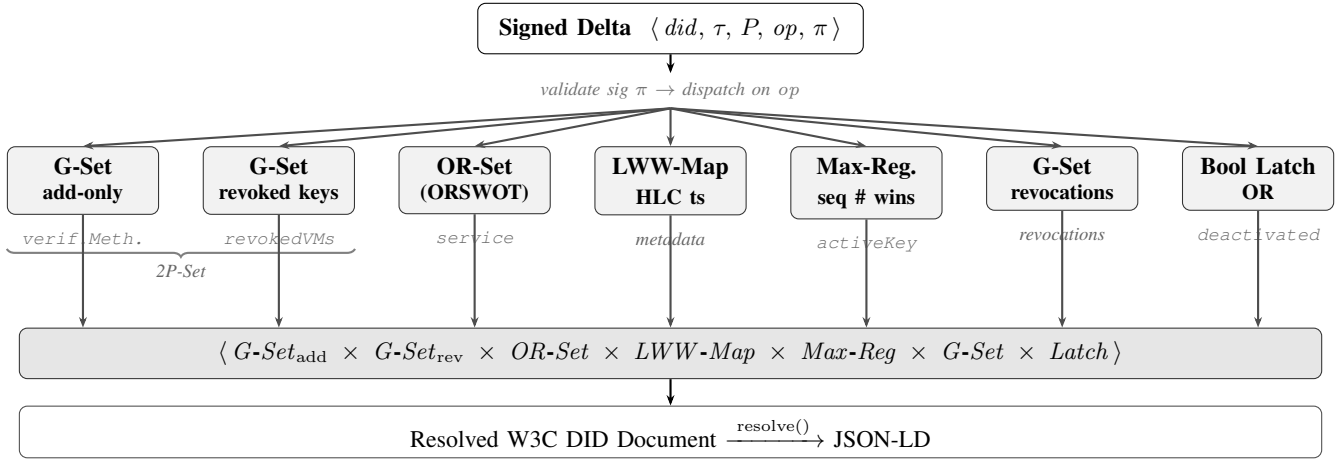
\begin{figure*}[t]
  \centering
  \begin{tikzpicture}[node distance=0.45cm and 0.2cm]
    \newlength{\figw}\setlength{\figw}{\textwidth}
    \node[delta] (delta)
      {\textbf{Signed Delta}\ \ $\langle\,\mathit{did},\,\tau,\,P,\,\mathit{op},\,\pi\,\rangle$};
    \node[below=0.25cm of delta, font=\scriptsize\itshape, text=black!55]
      (dispatch) {validate sig $\pi$ $\rightarrow$ dispatch on $\mathit{op}$};
    \node[crdt, below=0.5cm of dispatch, xshift=-0.4286\figw] (gset)
      {\shortstack{G-Set\\[1pt]\footnotesize add-only}};
    \node[crdt, below=0.5cm of dispatch, xshift=-0.2857\figw] (revvm)
      {\shortstack{G-Set\\[1pt]\footnotesize revoked keys}};
    \node[crdt, below=0.5cm of dispatch, xshift=-0.1429\figw] (orset)
      {\shortstack{OR-Set\\[1pt]\footnotesize (ORSWOT)}};
    \node[crdt, below=0.5cm of dispatch, xshift=0\figw] (lww)
      {\shortstack{LWW-Map\\[1pt]\footnotesize HLC ts}};
    \node[crdt, below=0.5cm of dispatch, xshift=0.1429\figw] (maxreg)
      {\shortstack{Max-Reg.\\[1pt]\footnotesize seq \# wins}};
    \node[crdt, below=0.5cm of dispatch, xshift=0.2857\figw] (revgset)
      {\shortstack{G-Set\\[1pt]\footnotesize revocations}};
    \node[crdt, below=0.5cm of dispatch, xshift=0.4286\figw] (latch)
      {\shortstack{Bool Latch\\[1pt]\footnotesize OR}};
    \node[role, below=0.1cm of gset]   {\texttt{verif.Meth.}};
    \node[role, below=0.1cm of revvm]  {\texttt{revokedVMs}};
    \node[role, below=0.1cm of orset]  {\texttt{service}};
    \node[role, below=0.1cm of lww]    {metadata};
    \node[role, below=0.1cm of maxreg] {\texttt{activeKey}};
    \node[role, below=0.1cm of revgset] {revocations};
    \node[role, below=0.1cm of latch]  {\texttt{deactivated}};
    \draw[decorate, decoration={brace, amplitude=3pt, mirror}, thick, black!50]
      ([yshift=-0.45cm]gset.south west) -- ([yshift=-0.45cm]revvm.south east)
      node[midway, below=2pt, font=\scriptsize\itshape, text=black!60] {2P-Set};
    \node[product, below=0.9cm of dispatch, yshift=-2.0cm,
          minimum width=0.95\figw] (product)
      {$\langle\,\mathit{G\text{-}Set}_{\mathrm{add}}\;\times\;\mathit{G\text{-}Set}_{\mathrm{rev}}\;\times\;\mathit{OR\text{-}Set}\;\times\;\mathit{LWW\text{-}Map}\;\times\;\mathit{Max\text{-}Reg}\;\times\;\mathit{G\text{-}Set}\;\times\;\mathit{Latch}\,\rangle$};
    \node[block, below=0.35cm of product,
          minimum width=0.95\figw] (doc)
      {Resolved W3C DID Document $\xrightarrow{\;\mathrm{resolve()}\;}$ JSON-LD};
    \draw[darrow] (delta.south) -- (dispatch.north);
    \foreach \n in {gset,revvm,orset,lww,maxreg,revgset,latch}{
      \draw[arrow] (dispatch.south) -- (\n.north);
    }
    \foreach \n in {gset,revvm,orset,lww,maxreg,revgset,latch}{
      \draw[arrow] (\n.south) -- (product.north -| \n.south);
    }
    \draw[darrow] (product.south) -- (doc.north);
  \end{tikzpicture}
  \caption{%
    \textbf{\didcrdt{} CRDT composition architecture.}
    An incoming signed delta $\langle\mathit{did},\tau,P,\mathit{op},\pi\rangle$
    is signature-verified and dispatched to the appropriate CRDT field.
    Verification methods and their revocations form a 2P-Set
    ($\mathit{authorized} = \mathit{added} \setminus \mathit{revoked}$).
    The product CRDT is projected to a W3C DID Core
    JSON-LD document by \texttt{resolve()}.
    By the CALM Theorem~\cite{calm}, the state-based merge is fully
    order-independent; the delta admission path is stateful and requires
    causal delivery for convergence.
    $\tau$: HLC triple $(\mathit{ms}, \mathit{ctr}, \mathit{node\_id})$; $\pi$: Ed25519 or secp256k1 signature.
    Each delta also commits to its causal parents $P$ by content hash (covered by $\pi$), so the history forms a Merkle DAG (Section~\ref{sec:delta}).
  }
  \label{fig:crdt-composition}
\end{figure*}

Figure~\ref{fig:crdt-composition} shows the architecture. The DID document state
is the \emph{product} of seven independent CRDT fields, each mapping a distinct
portion of the W3C DID Core model~\cite{w3c-did-core}; an incoming signed delta is
signature-verified and dispatched to the relevant field.

The product is again a CRDT~\cite{shapiro-crdts}, so the state-merge path stays
confluent and order-independent while the signed-delta path adds order-sensitive
authorisation (Section~\ref{sec:delta}).
Each field's merge semantics match the intended DID behaviour:

\begin{itemize}\setlength{\itemsep}{1pt}
  \item \textbf{Verification methods: 2P-Set}
        ($\mathit{authorized}=\mathit{added}\setminus\mathit{revoked}$, two
        grow-only G-Sets): a key is addable and permanently revocable but never
        resurrected, with full key history retained for audit. Any non-revoked key
        may authorise deltas.
  \item \textbf{Services: OR-Set (ORSWOT)}~\cite{orswot}: add-wins, so a re-added
        endpoint survives a concurrent stale removal without unbounded tombstones.
  \item \textbf{Metadata: LWW-Map}: last-writer-wins under an HLC
        timestamp~\cite{hlc}, with the public-key-derived $\mathit{node\_id}$ as a
        key-bound tiebreaker.
  \item \textbf{Key rotation: Max-Register}: highest sequence wins; it names the
        active controller for external authentication but does not gate delta
        authorisation.
  \item \textbf{Credential revocations: G-Set}: permanent and additive; never
        un-revokes.
  \item \textbf{Deactivation: boolean latch} (merge = OR): irreversible.
\end{itemize}

\subsection{Delta Model}
\label{sec:delta}

All mutations are \emph{signed deltas}: a tuple
$\langle\mathit{did}, \tau, P, \mathit{op}, \pi\rangle$ where $\tau$ is an HLC
triple $(\mathit{physical\_ms}, \mathit{counter}, \mathit{node\_id})$,
$P$ is the (sorted, deduplicated) set of content hashes of the delta's causal
parents (the frontier its signer had observed), so a DID's delta history forms
a \emph{Merkle DAG} (the genesis delta has $P=\varnothing$),
$\mathit{op}$ is one of eight typed operations
(\texttt{AddVerificationMethod}, \texttt{RevokeVerificationMethod},
\texttt{AddServiceEndpoint}, \texttt{RemoveServiceEndpoint},
\texttt{SetDocumentData}, \texttt{RotateKey}, \texttt{RevokeCredential},
\texttt{Deactivate}), and $\pi$ is an Ed25519 or secp256k1 ECDSA signature over
\texttt{canonical\_json(\{did,$\tau$,P,op\})}, so a delta cannot be re-parented
without invalidating its signature (\texttt{ed25519-dalek} and \texttt{k256} back
the two suites).
Each \texttt{AddVerificationMethod} carries a \texttt{SuiteType} discriminant and a
\texttt{relationships} list (which of the five W3C verification relationships the
key participates in; default \texttt{[authentication]}).
The $\mathit{node\_id}$ in $\tau$ MUST equal the lower 8 bytes of
BLAKE3(public key bytes), binding the tiebreaker to the signer's identity; the
validator recomputes it after signature verification and rejects on mismatch.

Validating an incoming delta is a five-step gate (well-formed, correctly signed,
prerequisites present, signer allowed, then merge), returning a \emph{three-valued}
result (\texttt{Unknown}/\texttt{Valid}/\texttt{Invalid}):
(1)~deserialise (deltas exceeding 64\,KiB are rejected before processing) and
idempotent dedup by content hash (a delta already held is a true no-op);
(2)~signature verification over $\langle\mathit{did},\tau,P,\mathit{op}\rangle$:
the signature MUST verify under a key present in the G-Set (for the genesis
delta, verification uses the embedded public key);
(3)~\emph{causal admission}: if the parents $P$ are not all present, the result
is \texttt{Unknown} and the delta is \emph{held pending} retry rather than
rejected;
(4)~\emph{authorisation} against the delta's \emph{causal past}, not current
state: the signer MUST be added there, and neither revoked nor the document
deactivated there. Concurrent deltas that a revocation or deactivation has not
yet reached are therefore both admitted; a current-state floor rejects only facts
imported via state-merge without their originating deltas;
(5)~CRDT merge (safe to repeat and reorder by CAI).

\smallskip\noindent
\textbf{Order-sensitivity caveat.}
The stateful admission pipeline is itself monotone: its three-valued result is a
flat lattice where \texttt{Unknown} resolves upward to \texttt{Valid}/\texttt{Invalid}
as a delta's causal closure completes and never reverses (an out-of-order delta is
held pending, not rejected). The protocol thus requires causal \emph{ordering} (a delta is applied after its
parents) but not a causal \emph{transport}: it reconstructs that order from the
Merkle-DAG parents, holding out-of-order deltas pending, so any delivery order
converges.

\subsection{CRUD Operations}

\textbf{Create:} derive the DID via genesis hash, sign and apply the genesis
delta; usable offline with no acknowledgement.
\textbf{Read:} project the product state to W3C DID Core JSON-LD; the
\texttt{versionId} is a BLAKE3 hash of observable state, and resolution does not
replay the log, so its cost is history-independent.
\textbf{Update:} wrap the operation in a signed delta with a fresh HLC timestamp
and broadcast; it converges once causally-prior deltas arrive, duplicates idempotent.
\textbf{Deactivate:} set the boolean latch; it propagates by OR, and per DID Core
\S8.2.1 the resolver then returns \texttt{didDocument:~null}.

\subsection{Concurrent Merge Semantics}

Concurrent edits resolve with no inter-replica communication: a concurrent
add/remove of one endpoint is add-wins (ORSWOT); equal-sequence
\texttt{RotateKey}s break by greater BLAKE3(\texttt{key\_ref}); same-key metadata
writes by greater HLC then $\mathit{node\_id}$; and \texttt{Deactivate} beats any
concurrent update (latch OR).

\section{Security Analysis}
\label{sec:security}

Against the adversary of Section~\ref{sec:sysmodel}:

\textbf{Signature-chain integrity.}
Every delta is signed over canonical JSON including the target DID and HLC
timestamp, preventing cross-DID replay and tiebreaker spoofing (the
$\mathit{node\_id}$ binding of Section~\ref{sec:delta}). Any non-revoked key
($\mathit{added}\setminus\mathit{revoked}$) may authorise new deltas; the 2P-Set
enables revocation while preserving convergence (a compromised key is permanently
excluded via the revocation G-Set), and since both halves are grow-only and
revocation irreversible, CAI holds.

\textbf{Key-compromise recovery (a fundamental limit).}
Authorisation is flat: \emph{any} non-revoked key may sign \emph{any} operation, so
a compromised key can add attacker keys, rotate, rewrite state, or deactivate the
DID before detection. Revoking it halts \emph{future} misuse but cannot undo what it
authorised (grow-only sets retain attacker keys; the deactivation latch is
irreversible). This exposure is no worse than single-controller ledger methods like
\texttt{did:ethr}~\cite{did-ethr} but weaker than Sidetree/\texttt{did:ion}~\cite{sidetree},
which separates update and recovery keys; \didcrdt{} has no such tier and, lacking a
consensus total order, no canonical state to override, so a compromised key's writes
\emph{merge} rather than being adjudicated away. It thus gives \emph{containment, not
recovery}; the future-work mitigation (Section~\ref{sec:conclusion}) closes this gap
via $M$-of-$N$ signatures for sensitive operations or a recovery-key policy.

\textbf{Metadata-ordering integrity.}
LWW-Map writes order by the HLC physical-millisecond component (the
$\mathit{node\_id}$ secures only the tiebreaker), so a Byzantine \emph{authorised}
signer can inflate its timestamp to dominate a metadata key; the HLC skew bound
limits this within the skew window. Metadata needing tamper-evident concurrency
should use the OR-Set field, whose add-wins merge keeps all concurrent writes.

\textbf{Genesis, replay, and DoS.}
Only \texttt{AddVerificationMethod} is permitted on an empty document, blocking
pre-genesis \texttt{Deactivate}/\texttt{RotateKey} attacks; forging the DID
$=$ BLAKE3-256(genesis) requires breaking 256-bit pre-image resistance. Replay is harmless by
idempotence (a delta already in the DAG is a no-op), and concurrent rotations are
all admitted and resolved by the Max-Register (higher sequence wins, BLAKE3
tiebreak). A 64\,KiB per-delta limit bounds per-delta cost, and resolution projects
materialised state independent of history length; bounding cumulative log growth
via checkpoint compaction is future work.

\textbf{Byzantine fault tolerance (delta path).}
\didcrdt{} achieves BFT on the delta path via content-addressed hashes,
deterministic validity (non-revoked G-Set membership), and application-layer
signatures (following Kleppmann~\cite{kleppmann-byzantine}): a Byzantine node
cannot forge valid deltas, and replay is idempotent. Each delta commits to its
causal past by hash, so that past is \emph{tamper-evident}: altering or dropping an
ancestor leaves a dangling parent hash the receiver re-requests, and equivocation
creates content-addressed forks that surface on reconcile. Withholding an unseen
concurrent branch stays undetectable (the DAG attests integrity, not uniqueness)
but only delays convergence; it cannot corrupt honest state.

\textbf{State-sync caveat.}
Convergence holds on every path; \emph{authenticity} (each delta signed by an
authorised key) only on the authenticated paths: the live gossip ingest, which
signature-verifies every inbound delta, and \texttt{merge\_verified\_bundle},
which re-derives state by replaying a content-addressed bundle of signed deltas
through full signature and admission checks. The unauthenticated
\texttt{merge\_state}, which imports another replica's state \emph{without}
re-verifying signatures, is retained only as a trusted in-process optimisation
(e.g.\ replica duplication within one operator's deployment); invoking it with an
untrusted source voids Byzantine safety. A remove cancels exactly the adds in its
causal past (fixed by its content-addressed parents), so delta replay resolves
concurrent add/remove add-wins like the state join ($\sqcup$).

\textbf{Sybil resistance.}
Minting a validly-formed DID costs only a keygen and one hash, so an adversary
could flood spurious creations. The planned (unimplemented) layered defence is a
20-bit genesis proof-of-work (${\sim}0.5$\,s per DID, ${\sim}10^6\times$ per-DID
batch-flood cost; updates exempt), per-IP creation rate limits, and optional
invitation codes for closed namespaces. Meanwhile the admission control of
Section~\ref{sec:resolution} (a node stores an unknown DID only when a local
resolution request solicits it) discards floods without storage.

\textbf{Discovery-layer availability.}
The discovery keypair is derived from the public DID
(Section~\ref{sec:resolution}), so anyone knowing a DID can publish over the
record. \emph{Redirection} cannot forge a document (genesis and delta-chain
verification authenticate content), costing only a wasted connection;
\emph{erasure} (overwriting the single-writer record) can suppress cold-start
discovery, a publish-rate race. Both touch first-contact only (nodes already
holding or gossiping the document are unaffected) and never risk integrity;
controller-signed records and erasure-resistant rendezvous are future work (the
race is inherent to identifier-derived keys, the model BitTorrent runs at scale).
Revocation is likewise irreversible by design.

\section{Resolution \& Discovery}
\label{sec:resolution}

Resolution (DID string to DID document) is a two-layer process.

\subsection{Local Resolution}

The \texttt{resolve()} function is a \emph{pure projection} requiring no network
call: it materialises the product CRDT state into the three-part envelope
mandated by DID Core \S7.1 (\texttt{didResolutionMetadata}, \texttt{didDocument},
\texttt{didDocumentMetadata}). Each CRDT field maps to its DID Core property: the
2P-Set yields \texttt{verificationMethod} (excluding revoked keys) and populates the
five verification-relationship arrays; the OR-Set yields \texttt{service}; the
LWW-Map yields document properties; the latch sets \texttt{deactivated}. Metadata
carries \texttt{created}/\texttt{updated} and a \texttt{versionId} (BLAKE3 over
observable state). For deactivated DIDs the resolver returns
\texttt{didDocument:~null} per \S8.2.1.

\subsection{Delta Discovery}

The pure core is deliberately transport-agnostic: any mechanism that
delivers signed deltas to a replica is sufficient. The core implements
transport-agnostic \emph{anti-entropy}: replicas exchange \emph{frontiers} (hashes
of the latest deltas each has seen) and ship only the missing deltas with their
ancestors, so cost is proportional to the difference, not the history. Three
discovery modes exist at varying maturity:

\textbf{(1)~Gossip mesh} (\emph{implemented}): the \texttt{sync} feature's
iroh-gossip~\cite{iroh} engine does frontier exchange (ANNOUNCE, REQUEST with
frontier, return the deltas above it); inbound deltas are signature-verified at the
trust boundary, and an integration test reconciles \emph{two real iroh endpoints}
over the wire.
\textbf{(2)~HTTP resolver} (\emph{implemented}): the \texttt{service} feature
(independent of \texttt{sync}) exposes an axum \texttt{GET /\{did\}} endpoint for
web-compatible resolution from an in-memory store.
\textbf{(3)~Out-of-band transfer} (\emph{core API}): for air-gapped use, signed
deltas (or, with the Section~\ref{sec:security} caveats, serialised state) transfer
via QR, NFC, Bluetooth, or USB.

Cold-start resolution (locating a never-seen DID's deltas) derives a keypair
deterministically from the DID's BLAKE3 hash: any holder publishes a signed pkarr
record~\cite{pkarr} (public-key addressable records over the mainline BitTorrent
DHT or an HTTP relay) advertising its iroh address, and a resolver derives the same
key, finds a holder, and bootstraps via empty-frontier anti-entropy; a forged record
only wastes a connection, never forging a document (genesis re-derivation
authenticates content). It is implemented and exercised by in-process and
cross-process (two service binaries against a pkarr relay stub) cold-start tests;
live public-DHT validation is future work.

\subsection{Cost Analysis}

Per-DID cost contrasts sharply with ledgers (protocol-level,
from~\cite{satybaldy-benchmarks} where available): per update \texttt{did:ethr}
costs \$0.066 at 12.9\,s and \texttt{did:hedera} \$0.00015 at 4.2\,s, whereas
\didcrdt{} has \emph{zero} fees and ${<}90\,\mu$s local merge (hosting \$4--8/mo
versus \$30--100). \didcrdt{}'s figures are local and exclude its not-yet-measured
availability layer, whereas ledger latency folds in global propagation and
consensus; a wide-area comparison awaits a complete stack.

Each signed delta is 200--500 bytes and the full log is retained, so a DID's
footprint is its materialised state plus history. Measured state RAM (log excluded)
is 1.0\,KiB (1~key/1~service), 4.4\,KiB (10/10), and 37.9\,KiB (100/100), so a
1\,GiB node holds ${\sim}$1.0\,M small DIDs' state; history adds 200--500 bytes per
update until checkpoint compaction (future work) bounds it.

\section{Implementation \& Evaluation}
\label{sec:impl}

\subsection{Reference Implementation}

The reference implementation is a Rust library (\texttt{did-crdt}, Rust $\geq$ 1.75)
with \textbf{zero \texttt{unsafe} blocks}, compiling to WebAssembly, in three feature
layers: \texttt{default} (pure core, no I/O), \texttt{sync} (iroh P2P delta gossip),
and \texttt{service} (axum HTTP resolver, independent of \texttt{sync}).
The pure core (${\sim}3{,}700$ non-blank lines) is self-contained and embeddable in
mobile (FFI), WASM, and edge contexts. Key dependencies: \texttt{crdts}~7
(G-Set, LWW-Register), \texttt{blake3}~1, \texttt{ed25519-dalek}~2, \texttt{serde};
the service-endpoint ORSWOT is implemented directly so a remove cancels exactly the
adds in its causal past.

\subsection{Correctness Testing}

The test suite comprises 262 functional tests (276 with \texttt{service}, 314 with
\texttt{sync}), most verifying distributed-system properties.
\emph{Property-based} tests (\texttt{proptest}, 256 cases each) check the CRDT laws
(CAI) over random delta sequences, plus rotation convergence at equal sequence,
revocation monotonicity, deactivation irreversibility, and signed-delta/state-join
agreement on concurrent add/remove of a service endpoint. \emph{Convergence} tests
simulate partition and reunion (50/30 concurrent deltas merging identically; all 24
permutations of four deltas byte-identical; three-replica order-independence).
Further suites cover the gossip state machine and HTTP CRUD, each CRDT primitive,
${\sim}$4\,M-iteration fuzzing of deserialisation (no panics), and Merkle-DAG
admission and authenticated sync, including a two-endpoint live-transport
convergence test and a forged-signature rejection on the gossip ingest.

\subsection{Performance}

Median merge latencies (Criterion~0.5, release$+$LTO, Apple M2 3.49\,GHz, via the
committed \texttt{benches/} harnesses) are 5--8\,$\mu$s on small documents
(1~key/1~service), 11--14\,$\mu$s medium (10/10), and 70--88\,$\mu$s large (100/100),
including Merkle-DAG admission but not signature verification; every small-document
operation exceeds 140\,K\,ops/s. Resolve latency is 1.8--74\,$\mu$s for the local
projection (no network round-trip, assuming the deltas are already held); the cost
of \emph{obtaining} them is the not-yet-measured availability layer
(Section~\ref{sec:resolution}), so this is not an end-to-end comparison with
blockchain reads.

\subsection{Deployment at Scale}

Garz{\'o}n et al.~\cite{garzon-did-6g} identify decentralised identity as critical
6G infrastructure. A key-rotating sensor fleet shows the gap: at ${\sim}$1{,}440
deltas/day per unit, a 10{,}000-sensor fleet on the cheapest ledger (XRPL,
\$0.000026/op~\cite{satybaldy-benchmarks}) costs \$137\,K/year with per-mutation
connectivity, whereas \didcrdt{} runs it offline fee-free; retained history
(${\sim}$3--7\,GB/day) needs the future-work compaction, and storage scales with
replicas, not a shared ledger.

\section{Discussion}
\label{sec:discussion}

\textbf{Resolution and \texttt{did:peer}.}
\didcrdt{} replaces the blockchain VDR with the CRDT state itself; in interactive
flows the owner presents $(\mathit{did}, [\delta_0, \ldots, \delta_n])$ directly,
like \texttt{did:peer}~\cite{did-peer} numalgo~2. But \texttt{did:peer} documents
are static and ephemeral, whereas \didcrdt{} merges multi-replica updates
deterministically, supports persistent identities with full key lifecycle, and
falls back to gossip/DHT resolution.

\textbf{W3C DID Core alignment.}
The resolver emits the mandatory \texttt{@context} and the DID Core \S7.1
resolution envelope, validated by construction and unit tests. We claim
\emph{alignment}, not certified compliance: the conformance suite and DID Method
Registry submission are future work.

\textbf{Applicability.}
\didcrdt{} suits owners controlling multiple intermittently-connected devices
(wallets, IoT fleets), frequent fee-free updates, or offline operation. It is less
suitable where an operation's legal effect depends on a single authoritative order
(e.g.\ eIDAS, where signature validity turns on revocation timing) or where the
identifier must be publicly discoverable with no prior relationship.

\textbf{Implementation limitations.}
\label{sec:limitations}
The core covers the full CRDT model; the gossip protocol, live iroh transport,
service-binary integration, and pkarr DHT cold-start discovery with admission
control are implemented and tested. Unimplemented: durable persistence (in-memory
store; a SQLite layer is specified but unbuilt); the sybil-resistance layers of
Section~\ref{sec:security}; hardened discovery; and wide-area convergence
measurement.

\section{Conclusion}
\label{sec:conclusion}

We have presented \didcrdt{}, a DID method aligned with W3C DID Core that achieves
consensus-free, offline-capable, cryptographically authorised identity management
by composing CRDTs. Mapping each document field to a CRDT makes the model a
join-semilattice, so CALM~\cite{calm} guarantees confluence without consensus given
eventual delivery, while application-layer signing separates \emph{authorisation}
from \emph{ordering}: the state-merge path needs no order, the signed-delta path
only causal delivery. We are explicit about the boundaries: Byzantine fault
tolerance holds for the signed-delta path but not the unauthenticated state-merge
API, and revocation gives containment, not recovery, from key compromise.

\textbf{Future work} spans stronger authorisation ($M$-of-$N$ threshold signing to
bound single-key compromise); compact state via Merkle-inclusion proofs plus
checkpoint compaction; completing the networking
stack (durable persistence, sybil-resistance layers, hardened discovery) with
wide-area measurement; W3C DID Method Registry submission and the conformance suite;
and formal verification of the merge laws.

\section*{Acknowledgments}

\textbf{AI Disclosure.} Per IEEE policy, the authors disclose that AI assistants
were used for drafting, implementation, and literature review; all work was
directed, reviewed, and validated by the authors, who take full responsibility.


\end{document}